\documentclass[11pt,fleqn]{article}
\usepackage{amsmath,amssymb,graphicx,epsfig}

\numberwithin{equation}{section}
\usepackage{color}

\usepackage{cancel}

\newcommand{\Tr}{\text{Tr}}

\newcommand{\re}{\text{Re}}

\begin{document}
\begin{flushright}
CERN-PH-TH-2015-268\\
\end{flushright}

\thispagestyle{empty}

\vspace{2cm}

\begin{center}
{\Large {\bf 
Gauging MSSM global symmetries and SUSY breaking in de Sitter vacuum
}}
\\
\medskip
\vspace{1.cm}
\textbf
{
I. Antoniadis$^{\,a,b}$, 
R. Knoops$^{\, c,d,e}$}
\bigskip

$^a$ {\small LPTHE, UMR CNRS 7589
Sorbonne Universit\'es, UPMC Paris 6, 75005 Paris France}

$^b$ {\small Albert Einstein Center, Institute for Theoretical Physics
Bern University, Sidlerstrasse 5, CH-3012 Bern, Switzerland }

$^c$ {\small CERN Theory Division, CH-1211 Geneva 23, Switzerland}

$^d$ {\small Section de Math\'ematiques, Universit\'e de Gen\`eve, CH-1211 Gen\`eve, Switzerland}

$^e$ {\small Instituut voor Theoretische Fysica, KU Leuven,
Celestijnenlaan 200D, B-3001 Leuven, Belgium}

\end{center}
\bigskip

\setcounter{page}{1}

\begin{abstract}
We elaborate on a recent study of a model of supersymmetry breaking we proposed recently, in the presence of a tunable positive cosmological constant, based on a gauged shift symmetry of a string modulus, external to the Standard Model (SM) sector. Here, we identify this symmetry with a global symmetry of the SM and work out the corresponding phenomenology. A particularly attracting possibility is to use a combination of Baryon and Lepton number that contains the known matter parity and guarantees 
absence of dimension-four and five operators that violate B and L.

\end{abstract}

\section{Introduction}

In a recent paper~\cite{RK3}, we performed a detailed study of the phenomenology of a supergravity model of supersymmetry breaking~\cite{Zwirner,RK1}, having a metastable de Sitter (dS) vacuum with a tiny (tunable) cosmological constant, independent of the supersymmetry breaking scale. The model is based on a shift symmetry associated to a string modulus (dual to a two-index tensor) that we consider to be in the dilaton supermultiplet for definiteness, which is gauged using a vector multiplet. Depending on the K\"ahler basis, the shift symmetry becomes a gauged R-symmetry that fixes the form of the superpotential as a single exponential and allows for the presence of a Fayet-Iliopoulos (FI) term. The model has thus three parameters: the strengths of the superpotential and the FI-term, and the exponent of the exponential superpotential. The first two can be tuned to fix the vacuum energy at a tiny positive value, while the latter determines the vacuum expectation value (VEV) of the dilaton. The overall scale then controls the supersymmetry breaking, or equivalently the gravitino mass, driven by expectation values of both F- and D-auxiliary components of the chiral and vector multiplet.

The model can be easily coupled to an observable sector containing a supersymmetric extension of the Standard Model (MSSM). To avoid anomalies~\cite{RK2}, we considered all MSSM fields inert under the shift symmetry, in a K\"ahler basis where the $U(1)$ is not an R-symmetry, and a constant (modulus independent) gauge kinetic function. In the simplest case however scalar masses are tachyonic which can be avoided, without modifying the main properties of the model, by introducing either a new `hidden sector' field participating in the supersymmetry breaking, similar to the Polonyi field~\cite{Polonyi}, or dilaton dependent matter kinetic terms~\cite{RK3}. In both cases, an extra parameter is introduced with a narrow range of values, in order to satisfy all required constraints.
All scalar soft masses and trilinear A-terms are generated at the tree-level and are universal under the assumption that matter kinetic terms are independent of the `Polonyi' field, while gaugino masses are generated at the quantum level, via the so-called anomaly mediation contributions~\cite{gauginomass}, and are naturally suppressed compared to the scalar masses.

It follows that the low energy spectrum is very particular and can be distinguished from other models of supersymmetry breaking and mediation, such as mSUGRA and mAMSB. It consists of light neutralinos, charginos and gluinos, where the experimental bounds on the (mostly bino-like) LSP and the gluino force the gravitino mass to be above 15 GeV and the squarks to be very heavy, with the exception of the stop which can be as light as 2 TeV.

In this work, we study the possibility that the gauged shift symmetry is identified with a known global symmetry of the Standard Model (SM), or more generally its supersymmetric extension, keeping the nice properties of the model, namely the existence of the metastable dS vacuum with a tunable cosmological constant and a viable spectrum of superparticles consistent with all experimental constraints. A particular attracting possibility is to use a symmetry that contains the usual R-parity, or matter-parity (depending on the K\"ahler basis) of the MSSM. We find that this is indeed possible and analyze explicitly the anomaly free symmetry $B-L$ (when adding three right-handed neutrinos), where $B$ and $L$ stand for the baryon and lepton number respectively, or the combination $3B-L$ which has the advantage of forbidding all dimension-four and dimension-five operators violating baryon or lepton number in MSSM. It turns out that the phenomenology of these two cases is similar to the results we found in the case where SM fields are inert under the shift symmetry~\cite{RK3}, with a few minor differences, such as that the stop squark can become lighter to about 1.5 TeV. Actually the model contains an extra parameter, the unit of $B-L$ charge $q$ for the SM fields, that allows to extrapolate between the present analysis and the one of Ref.~\cite{RK3}. It turns out though that $q$ is bounded from the requirement of existence of the electroweak vacuum.

We also address the question if the problem of tachyonic scalar masses can be solved without adding extra field or modifying the matter kinetic terms, by appropriately choosing the transformations of the MSSM fields, due to the extra D-term contribution in the scalar potential since SM fields are now charged under the $U(1)$. However, the answer turns out to be negative due to constraints arising from the existence of the usual electroweak vacuum. Finally, we analyze the phenomenological implications of the extra $U(1)$ and 
we find that its coupling is too small to have possible experimental signatures in colliders at present energies.

The outline of the paper is the following. In Section~2, we present a short review of our model and give our conventions. In Section~3, we analyze the possibility of identifying the gauged shift symmetry with the $B-L$; subsection~3.1 contains a short review of R-parity versus Matter-parity for self-completeness, while in subsection~3.2, we work out the model and its phenomenology; we also comment on the case of $3B-L$ but we don't repeat the analysis, since the results are very similar. In Section~4, we consider the most general global symmetry and address the question of tachyonic scalar masses without extra field or modification of the matter kinetic terms. Section~5 contains a brief summary of our results and the main conclusions. 
Finally, there are three appendices. Appendix~\ref{appendix:anomalies} contains the computation of anomalies, their cancellation and the one-loop corrections to the gaugino masses. 
In Appendix~\ref{Appendix:p2}, we study a possible leftover case from our past analyses, where the $U(1)$ gauge kinetic term is linear (allowed by the shift symmetry) and the coefficient of the logarithm in the dilaton K\"ahler potential is $p=2$ (instead of $p=1$), implied by the tunability of the cosmological constant~\cite{Zwirner,RK1}; we find that actually this is not a phenomenologically viable possibility.
In Appendix~\ref{Appendix:nonzerobeta} we verify the assumption made in section~\ref{sec:other} that the linear contribution to the gauge kinetic function is very small. 

\section{Conventions and review of the model} \label{sec:model}

In this section\footnote{Throughout this paper we use the conventions of \cite{VP}.} we review a class of metastable de Sitter vacua proposed in \cite{Zwirner,RK1} and further analyzed in \cite{RK3}, which have a tunable (infinitesimally small) value of the cosmological constant and a TeV gravitino mass. The minimal version of the model consists, in addition to the supergravity multiplet, of a chiral multiplet $S$ and a vector multiplet associated with a shift symmetry of the scalar component $s$ of the chiral multiplet
\begin{align} \delta s = -ic \theta. \label{shift} \end{align}
The K\"ahler potential, superpotential and gauge kinetic function are given by
\begin{align} \mathcal K &= - \kappa^{-2} \log (s + \bar s) + \kappa^{-2} b (s + \bar s), \notag \\ W &= \kappa^{-3} a , \notag \\ f(s) &= 1, \label{model:basic} \end{align} 
where $a,b$ and $c$ are constants which can be tuned to allow for an infinitesimally small cosmological constant and a TeV gravitino mass. The gauge kinetic function can in principle be any real constant $f(s) = \gamma$. However, as far as the minimization of the potential is concerned, this constant can be put to 1 by a rescaling of $c$. The scalar potential is given by
\begin{align} V &= V_F + V_D, \end{align} where the F-term contribution is given by  
\begin{align} V_F &= e^{\kappa^2 \mathcal K} \left( - 3 \kappa^2 W \bar W  + \nabla_\alpha W g^{\alpha \bar \beta} \bar \nabla_{\bar \beta} \bar W \right) \notag \\ &= \kappa^{-4} |a|^2 \frac{e^{b(s + \bar s)}}{s + \bar s} \sigma_s, && \sigma_s=  -3 + \left( b(s + \bar s) - 1 \right)^2 ,\end{align} 
and the D-term contribution to the scalar potential is given by \begin{align} V_D &= \frac{1}{2} \left( \re f \right)^{-1 \ AB} \mathcal P_A \mathcal P_B \notag \\ &= \kappa^{-4}   \frac{c^2}{2} \left( b - \frac{1}{s + \bar s} \right)^2. \end{align}
Here the indices $\alpha, \beta$ label the chiral multiplets, the indices $A,B$ indicate the different gauge groups and $\kappa$ is the inverse of the reduced Planck mass, $m_p = \kappa^{-1} = 2.4 \times 10^{15} $ TeV. The K\"ahler covariant derivative and moment maps are given by
  \begin{align}\nabla_\alpha W &= \partial_\alpha W(z) + \kappa^2 (\partial_\alpha \mathcal K) W(z), \notag \\    \mathcal P_A &= i(k_A^\alpha \partial_\alpha \mathcal K - r_A). \end{align}
The Fayet-Iliopoulos contributions $r_A$ are fixed by the relation \begin{align} W_\alpha k^\alpha_A = -\kappa^2 r_A W, \end{align} where $k^\alpha_A$ are the Killing vectors. 

For $b \geq 0$ this scalar potential always allows for an anti-de Sitter (AdS) minimum. We therefore focus on the case $b<0$, where it was shown in \cite{RK3} that this model allows for an infinitesimally small cosmological constant $\Lambda$ by tuning the parameters $a,b,c$ such that
\begin{align} b \langle s + \bar s \rangle &= \alpha \approx -0.233153, \notag \\ \frac{b c^2}{a^2} &= A(\alpha) + \frac{2 \kappa^4 \Lambda \alpha^2}{a^2 b (\alpha - 1)^2}, & A(\alpha) &=  2 e^\alpha \alpha \frac{ 3 - (\alpha -1)^2}{ (\alpha - 1)^2} \approx -0.359291, \label{model:tuning} \end{align} 
where $\alpha$ is the negative root of $-3 + (\alpha -1 )^2(2 - \alpha^2/2) = 0$ close to 0.23.  A problem arises when this model is used as a hidden sector for supersymmetry breaking that is then communicated to the MSSM via gravity mediation:  It turns out that the resulting soft scalar masses for the MSSM fields are tachyonic.  It was shown in \cite{RK3} that this problem can however be avoided by introducing an extra Polonyi-like field  $z$ (see eqs.~(\ref{model:polonyi})) or by allowing a non-canonical K\"ahler potential for the MSSM superfields (see eqs.~(\ref{model_noncan})), while maintaining the desirable properties of the scalar potential, such as an infinitesimally small cosmological constant and a separately tunable gravitino mass. 

The model including an extra Polonyi-like field has a K\"ahler potential, superpotential and gauge kinetic function given by
\begin{align} \mathcal K &= -\kappa^{-2} \log (s + \bar s) + \kappa^{-2} b (s + \bar s) + z \bar z, \notag \\ W &= \kappa^{-3} a (1+ \gamma \kappa z) , \notag \\ f(s) &= 1. \label{model:polonyi} \end{align}
The scalar potential is
\begin{align} V &= V_F + V_D, \notag \\ V_F &= \kappa^{-4} |a|^2 \frac{e^{b(s + \bar s) + \kappa^2 z \bar z}}{s + \bar s} \left( \sigma_s A(z,\bar z) + B(z, \bar z) \right),  \notag \\ V_D &= \kappa^{-4}   \frac{c^2}{2} \left( b - \frac{1}{s + \bar s} \right)^2, \label{model:polonyi-scalarpot} \end{align}
where \begin{align}  A(z, \bar z) &= \left| 1 + \gamma \kappa z \right|^2, \notag \\  B(z, \bar z) &= \left| \gamma + \kappa \bar z + \gamma \kappa^2 z \bar z \right|^2. \label{ABz}\end{align}
The role of the extra hidden sector field $z$ is to give a (positive) F-term contribution to the  scalar potential, which in turn gives a positive contribution to the soft mass squared for the MSSM-like fields at the cost of introducing an extra parameter $\gamma$ to the model. This parameter is however very constrained:
\begin{align}  \gamma \in \left] 0.5, 1.707 \right[, \end{align}
where the lower bound arises due to an instability of the potential when the imaginary part of $z$ acquires a VEV, and the upper bound arises from the requirement of the tunability of the scalar potential. When experimental constraints are taken into account, in particular the gluino mass lower limit, the lower bound on the parameter $\gamma$ rises to about $1.1$. 

A careful tuning of the parameters then allows us to obtain a tunably small and positive value of the minimum of the potential by
  \begin{align} \frac{c^2}{a^2 } =-2 \frac{\alpha}{b} e^{\alpha + t^2} \left[\frac{\sigma_s A(t) + B(t)}{(\alpha-1)^2} \right] + \frac{2 \alpha^2}{(\alpha-1)^2} \frac{\kappa^4 \Lambda}{a^2 b^2} , \label{polonyi:vacuum}\end{align}
where we focus on real $z=\bar z = \kappa^{-1} t$ and   \begin{align} A(t) &= (1+\gamma t)^2, \notag \\ B(t) &= (\gamma + t + \gamma t^2)^2 \ .  \label{ABt} \end{align} 
 For a given $\gamma$, only one free parameter remains in the model, which can be taken to be the gravitino mass $m_{3/2}$, given by
  \begin{align} m_{3/2} = \kappa^2 e^{\kappa^2 \mathcal K/2} W =  \kappa^{-1} a \sqrt { \frac{b}{ \alpha} } e^{\alpha/2 + t^2/2} \left( 1 + \gamma t \right).  \label{gravitino_polonyi} \end{align}
 The masses of the hidden sector particles (including the gauge boson of the extra $U(1)_R$)\footnote{Even though in the K\"ahler basis (\ref{polonyi:vacuum}) the shift symmetry is technically not an R-symmetry, we will continue to label it with the index $R$ throughout this paper.} turn out to be proportional to $m_{3/2}$. When this model is used as a hidden sector, where supersymmetry breaking is communicated to the MSSM via gravity mediation, the soft terms turn out to be (in the standard notation)
\begin{align}   m_0^2 &= m_{3/2}^2  \left[ \left( \sigma_s + 1 \right) + \frac{(\gamma + t + \gamma t^2)^2}{(1 + \gamma t)^2}\right], \notag \\   A_0&= m_{3/2} \left[  (\sigma_s +3)  + t \frac{ (\gamma + t +\gamma t^2) } {1+\gamma t} \right], \notag \\   B_0 &= m_{3/2}  \left[  (\sigma_s + 2) + t \frac{ (\gamma + t + \gamma t^2) }{(1+\gamma t)}  \right]. \label{softterms_polonyi}   \end{align}
The gaugino masses arise at one-loop \cite{gauginomass} and are given by (see eq.~(\ref{Appendix:gauginomassAM}))
\begin{align}   M_1 &= 11 \frac{g_Y^2}{16 \pi^2} m_{3/2} \left[ 1 - (\alpha -1)^2 -  \frac{ t(\gamma + t + \gamma t^2)}{1 + \gamma t}  \right], \notag \\   M_2 &=  \frac{g_2^2}{16 \pi^2} m_{3/2} \left[1 - 5 (\alpha-1)^2 -5 \frac{ t (\gamma + t + \gamma t^2)}{1 + \gamma t} \right], \notag \\   M_3 &= - 3 \frac{g_3^2}{16 \pi^2} m_{3/2} \left[ 1 + (\alpha - 1)^2 + \frac{ t (\gamma + t + \gamma t^2) }{1 + \gamma t} \right]. \label{m1m2m3polonyi}   \end{align}  
It turns out \cite{RK3} that the low energy spectrum can be distinguished from minimal scenarios of supersymmetry breaking and mediation such as mSUGRA and mAMSB. 

Another possible solution to the negative scalar soft masses squared involves a non-canonical K\"ahler potential for the MSSM superfields, and the model is given by
  \begin{align} \mathcal K &= - \kappa^{-2} \log (s + \bar s) + \kappa^{-2} b (s + \bar s) + (s + \bar s)^{-\nu} \sum \varphi \bar \varphi , \notag \\ W &= \kappa^{-3} a + W_{MSSM} , \notag \\  f(s) &= 1, \ \ \ \ \ \ \ f_A(s)=1/g_A^2. \label{model_noncan} \end{align} 
Since the low energy phenomenology of this model is comparable with the one above with an extra parameter $\nu$ instead of $\gamma$, we do not further discuss this model.

In section \ref{sec:B-L} we investigate the effects of allowing a charge proportional to B-L for the MSSM superfields in the model with the extra Polonyi-like field, while in section \ref{sec:other} we show that a third possible solution to the tachyonic masses including a D-term contribution to the scalar soft masses squared does not contain viable solutions to the RGE (Renormalization Group Equations).
The problem with tachyonic masses can in principle also be solved by taking a linear gauge kinetic function $f(s)=s$ and $p=2$ in $\mathcal K = -\kappa^{-2} p \log(s+ \bar s)$. However, in the Appendix \ref{Appendix:p2} we show that this leads to an unacceptable high value of the gravitino mass.

\section{B-L case} \label{sec:B-L}
\subsection{R-parity versus Matter Parity}

In the context of MSSM, a global R-parity is usually imposed to forbid terms in the Lagrangian that may lead to proton decay. Although it is widely known \cite{SUSYprimer} that R-parity can be formulated in such a way that it is not an R-symmetry in the technical sense of the word, for self-contained presentation and benefit of the reader, we remind below of this fact.

The most general gauge-invariant and renormalizable superpotential for the MSSM would not only include the usual terms 
  \begin{align} W_{MSSM}= y_u^{ij} \bar u_i Q_j \cdot H_u - y_d^{ij} \bar d_i Q_j \cdot H_d - y_e^{ij} \bar e_i L_j \cdot H_d + \mu H_u \cdot H_d, \label{MSSMsuperpot} \end{align}
 but also the following baryon- and lepton-number violating interactions
\begin{align} W_{\Delta L =1} &= \frac{1}{2} \lambda^{ijk} L_i L_j \bar e_k + \lambda'^{ijk } L_i Q_j \bar d_k + \mu'^i L_i H_u, \notag \\ W_{\Delta B =1} &= \frac{1}{2} \lambda''^{ijk}\bar u_i \bar d_j \bar d_k. \label{deltabl} \end{align}
The chiral superfields carry baryon number $B = 1/3$ for $Q_i$, $B=-1/3$ for $\bar u_i, \bar d_i$ and $B=0$ for all the others. Similarly, $L_i$ and $\bar e_i$ carry lepton number $+1$ and $-1$, respectively, while all other superfields have vanishing lepton number. Since the baryon and lepton number violating processes (\ref{deltabl}) are not seen experimentally, these terms should be absent (or sufficiently suppressed). This is usually done by imposing that a discrete R-parity is preserved. The R-parity of a field is given by
\begin{align} P_R = (-1)^{3(B-L) + 2s},\end{align}
with $s$ its spin. Note that the Standard Model particles and Higgs bosons carry $P_R = +1$, while the 'sparticles' (squarks, sleptons, gauginos and Higgsinos) have $P_R =-1$. Also, since every interaction vertex contains an even number of $P_R = -1$ particles, this implies that the lightest sparticle (LSP) with $P_R=-1$ must be absolutely stable. If this LSP interacts only weakly with ordinary matter it can be an excellent dark matter candidate. Note also that since the different fields of the same multiplet carry a different R-parity, this symmetry does not commute with supersymmetry. 

Although this assignment appears to be quite natural in a supersymmetric context, it should be stressed that one can equivalently forbid the terms (\ref{deltabl}) by imposing conservation of matter parity \cite{matterparity}. The matter parity $P_M$ of a superfield (as opposed to R-parity, which is defined separately on each component field) is defined as
\begin{align} P_M = (-1)^{3(B-L)}. \label{matterparity} \end{align}
Note that since the matter parity of all fields within a given supermultiplet is the same, this symmetry does commute with supersymmetry. Since for the scalar components ($s=0$) the matter parity is the same as the R-parity, it is completely equivalent to impose either matter parity or R-parity as a symmetry on the theory. Moreover, R-parity and matter parity only differ by the fermion number, which is an exact parity symmetry by itself.

We conclude that imposing matter parity or R-parity is completely equivalent. While the R-parity interpretation can be useful to easily abstract its phenomenological consequences, from a model building point of view it is far more natural to impose matter parity (\ref{matterparity}), since (in contrast with R-parity) it commutes with supersymmetry. In fact, since R-parity is equivalent to the (non-R) matter parity, this shows that there is nothing intrinsically 'R' about R-parity.

Matter parity is nothing else but $3(B-L)$, it therefore seems an obvious choice in our context to see whether one can consistently use it as the $U(1)$ symmetry we need in our toy model of supersymmetry breaking, by giving a charge proportional to $B-L$ to the MSSM superfields, as to exclude also the terms (\ref{deltabl}) from the superpotential and thus taking over the role of R-parity. 

It should however be noted that in principle one can also have dimension 5 operators that violate Baryon and/or Lepton number (see for example \cite{dim5} and references therein)
\begin{align}  W_{\text{dim 5}} = & \kappa^{(0)}_{ij} H_u L_i H_u L_j + \kappa^{(1)}_{ijkl}Q_i Q_j Q_k L_l + \kappa^{(2)}_{ijkl} \bar u_i \bar u_j \bar d_k \bar e_l \notag \\  &+ \kappa^{(3)}_{ijk} Q_i Q_j Q_k H_d + \kappa^{(4)}_{ijk} Q_i \bar U_j \bar e_k H_d + \kappa^{(5)}_{i} L_i H_u H_u H_d. \label{dim 5} \end{align}
Here the various couplings $\kappa^{(n)}$ have inverse mass dimensions and can be generated by a high-energy microscopic theory, such as a supersymmetric grand unified theory or string theory.
While R-parity forbids the terms in the last line of eq.~(\ref{dim 5}), all terms in the first line are still allowed. Imposing a B-L symmetry additionally forbids the terms proportional to $\kappa^{(0)}_{ij}$. The terms proportional to $\kappa^{(1)}_{ijkl}$ and $\kappa^{(2)}_{ijkl}$ are still allowed.  It should however be noted that a $3B-L$ symmetry (which has the same parity $(-1)^{3B-L} = (-1)^{3B-3L}$ on MSSM fields) forbids all the above dimension 5 operators.  However, in contrast with a gauged $B-L$ which can be made anomaly-free upon the inclusion of three right-handed neutrinos,  a gauged $3B-L$ contains a cubic $U(1)_{3B-L}^3$, and mixed $U(1)_{3B-L} \times SU(2)$ and $U(1)_{3B-L} \times U(1)_Y^2$ anomalies which should be canceled by a Green-Schwarz mechanism. 

\subsection{The model extended with B-L charges}

As discussed above, we now explore the possibility to give the MSSM superfields (denoted by $\varphi_i$) a charge $q_i$ under the extra $U(1)_R$, proportional to $B-L$, extending the model~(\ref{model:polonyi}).  This means that $Q, \bar u$ and $\bar d$ have charges $q/3, -q/3$ and $-q/3$, respectively. The Higgs superfields do not carry a charge and the leptons $L$ and $\bar e$ carry a charge $-q$ and $+q$ respectively.

First, this gives contributions to the D-term part of the scalar potential, and one should check that this does not ruin its stability. The scalar potential is now given by
\begin{align}  V &= V_F + V_D, \notag \\  V_D &= \frac{1}{2} \left( -\frac{\kappa^{-2} c}{s + \bar s}  + \kappa^{-2} bc - \sum q_i \varphi_i \bar \varphi_i \right)^2, \label{polonyi:scalarpot2} \end{align}
where $V_F$ is the same as in eq.~(\ref{model:polonyi-scalarpot}). The D-term part will give an extra contribution to the soft scalar masses of the matter fields $\varphi_i$. The restriction that these remain non-tachyonic gives
\begin{align}  0< \left. \partial_{\varphi_i} \partial_{\bar \varphi_i}V \right|_{\varphi = 0}  = \kappa^{-2} a^2 \frac{e^{b (s + \bar s) + t^2} }{s + \bar s} \left(A(t) (\sigma_s + 1) + B(t) \right) + \kappa^{-2} q_i c \left(\frac{1}{s + \bar s} - b \right),\notag \end{align}
where $A(t)$ and $B(t)$ are given in eqs.~(\ref{ABt}). Since $q_i$ can be either positive or negative, and the first term on the r.h.s. is positive for the VEVs of $t$ and $s$ (see \cite{RK3}), it follows that
\begin{align}  |q| < \frac{a^2}{c} e^{\alpha + t^2} \frac{A(t) (\sigma_s+1) + B(t) }{1-\alpha},  \end{align}
which can be rewritten as (by use of eqs.~(\ref{polonyi:vacuum}) and (\ref{gravitino_polonyi}))
\begin{align}  |q| < q_{\text{max}} = \kappa m_{3/2} \frac{A(t) (\sigma_s+1) + B(t)}{\sqrt{|A(t) \sigma_s + B(t)|}} \frac{1}{\sqrt{2} (1 + \gamma t) }.  \label{qconstraint} \end{align}
However, we will show below that one actually needs $|q|/q_{\text{max}}< 0.013$ in order to find a viable solution to the RGE.
Note that the constraint~(\ref{qconstraint}) can be rewritten  (by using the relation~(\ref{polonyi:vacuum})) as
\begin{align}   |q| < q_{\text{max}} = bc \ \frac{A(t)(\sigma_s +1) + B(t)}{A(t) \sigma_s + B(t)} \frac{1-\alpha}{\alpha}, \end{align} 
where $\kappa^{-2} bc$ is the Fayet-Iliopoulos constant in the scalar potential~(\ref{polonyi:scalarpot2}).

While the mixed $U(1)_R \times U(1)_Y^2$,  $U(1)_R \times SU(2)$ and  $U(1)_R \times SU(3)$ anomalies vanish, the cubic anomaly vanishes only upon the inclusion of three right-handed neutrinos which are singlets under the Standard Model gauge groups. Otherwise, the cubic anomaly is proportional to
\begin{align}  \text{Tr} Q^3 = -3q^3, \end{align}
but the mixed anomalies still vanish. In this case, the cubic anomaly should be canceled by a Green-Schwarz counter term (see appendix \ref{appendix:anomalies}), provided
\begin{align}   f(s) &= 1 + \beta_R s, \notag \\  \beta_R &= - \frac{q^3}{4\pi^2 c}. \end{align} 

The gaugino masses are generated at one loop from anomaly mediaton, given by eqs.~(\ref{m1m2m3polonyi}), while the other soft supersymmetry breaking terms are given by
\begin{align}   m_{0,i}^2 &= m_{3/2}^2  \left[ \left( \sigma_s + 1 \right) + \frac{(\gamma + t + \gamma t^2)^2}{(1 + \gamma t)^2}\right] + \kappa^{-2} q_i bc \left(\frac{1}{\alpha} - 1 \right) , \notag \\   A_0&= m_{3/2} \left[  (\sigma_s +3)  + t \frac{ (\gamma + t +\gamma t^2) } {1+\gamma t} \right], \notag \\   B_0 &= m_{3/2}  \left[  (\sigma_s + 2) + t \frac{ (\gamma + t + \gamma t^2) }{(1+\gamma t)}  \right].    \end{align}
Or, by using
\begin{align} bc = m_{3/2} \kappa \frac{ \sqrt{-2\left( A(t) \sigma_s + B(t) \right)} }{1+ \gamma t} \frac{\alpha}{1-\alpha}, \label{bc}\end{align}
the soft terms can be written as
\begin{align}   m_{0,i}^2 &= m_{3/2}^2  \left[ \left( \sigma_s + 1 \right) + \frac{(\gamma + t + \gamma t^2)^2}{(1 + \gamma t)^2}\right] + \kappa^{-1} m_{3/2} q_i \frac{\sqrt{-2 \left(A(t) \sigma_s + B(t) \right)}}{1+\gamma t} , \notag \\   A_0&= m_{3/2} \left[  (\sigma_s +3)  + t \frac{ (\gamma + t +\gamma t^2) } {1+\gamma t} \right], \notag \\   B_0 &= m_{3/2}  \left[  (\sigma_s + 2) + t \frac{ (\gamma + t + \gamma t^2) }{(1+\gamma t)}  \right]. \label{softtermsB-L}   \end{align}
Note that the relation \begin{align} A_0 = B_0 + m_{3/2} \label{AB-relation}\end{align} still holds, as in \cite{RK3}.

In \cite{RK3} the special case $q=0$ was analyzed in full detail; it was shown that for $\gamma < 1.1$ no solutions to the RGEs exist that satisfy eq.~(\ref{AB-relation}). 
Moreover, it was shown that for $\gamma \rightarrow 1.1$ the mass of the lightest stop $m_{\tilde t}$ can become very small. By imposing a lower bound of about $m_{3/2}\geq 15$ TeV on the gravitino mass, which originates from a lower bound of about 1 TeV on the gluino mass, it was shown that the mass of the lightest stop can be as low as about 2 TeV, while the masses of the other squarks remain high ($>10$ TeV).

As it turns out, the only considerable effect of a nonzero charge $q$ to the sparticle spectrum is for the lightest stop, whose dependence on the input parameter $q/q_{\text{max}}$ for $m_{3/2}=15 \text{ TeV}$ and $\gamma=1.1$ is plotted in figure 1 \cite{softsusy}.
\begin{figure} \centering \includegraphics[width=0.5\textwidth]{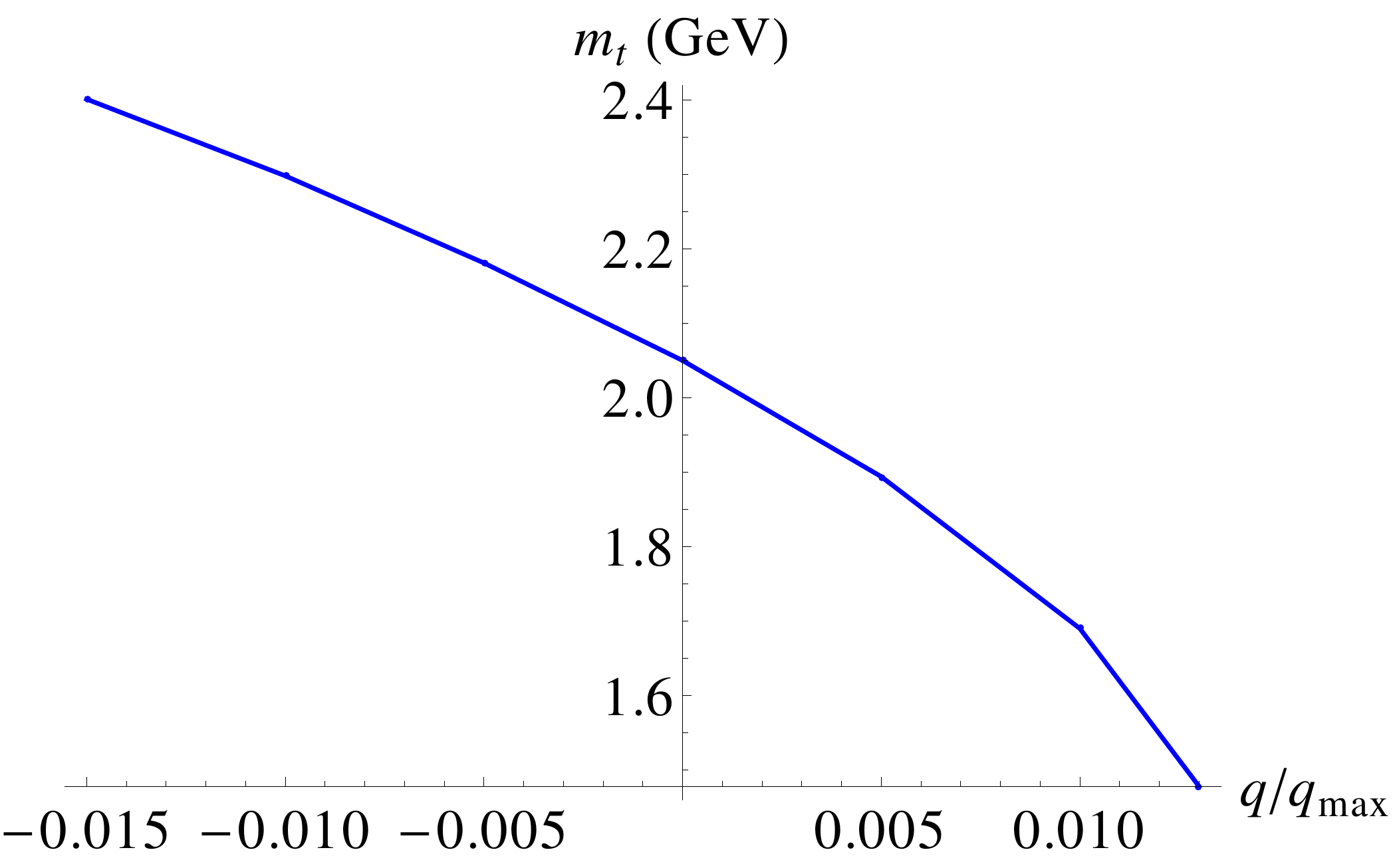} \label{B-L_mt_qqmax}  \caption{The mass of the lightest stop squark as a function of the charge $q/q_{\text{max}}$ for $\gamma=1.1$ and $m_{3/2}=15 \text{ TeV}$. The gravitino mass is chosen so that the gluino mass is right above the experimental bound of 1 TeV (while other experimental bounds such as the neutralino and charginos are also satisfied). A positive $q$ corresponds to the scalar soft masses $m_Q$ and $m_e$ being heavier than $m_L$ and $m_d = m_u$ (see eq.~(\ref{softtermsB-L})). For $q/q_{\text{max}} > 0.013$ no solutions to the RGE were found.} \end{figure} 
For $q/q_{\text{max}} > 0.013$, no solutions to the RGE were found. A lower limit for the mass of the lightest stop of about $1.5$ TeV is found when $q/q_{\text{max}} \rightarrow 0.013$. 

It should however be noted that, since anomalies are canceled by a Green-Schwarz mechanism, one can in principle choose different charge allocations for the MSSM fields which allow the terms in the superpotential (\ref{MSSMsuperpot}), while forbidding the Baryon and Lepton violating terms (\ref{deltabl}) and the dimension five operators (\ref{dim 5}).
As mentioned above, a gauged $B-L$ forbids the terms in eq.~(\ref{deltabl}), but it still allows certain dimension five operators. 
This can be solved by gauging $3B-L$. A gauged $3B-L$ is anomalous and its $U(1)_{3B-L} \times U(1)_Y^2$ and $U(1)_{3B-L} \times SU(2)$ anomalies are proportional to
\begin{align} \mathcal C_1 &= -3q, \notag \\ \mathcal C_2 &= 6q,\end{align}
while the $U(1)_{3B-L}^2 \times U(1)_Y$ and $U(1)_{3B-L} \times SU(3)$ anomalies vanish.
As we outline in the Appendix \ref{appendix:anomalies}, this results in a contribution to the gaugino masses eq.~(\ref{gauginomassGS}) given by
\begin{align}
 M_1 &= - g_Y^2 \frac{3 \alpha (\alpha-1)}{8 \pi^2} \frac{q}{bc}{m_{3/2}}, \notag \\
 M_2 &=  g_2^2 \frac{6 \alpha (\alpha-1)}{8 \pi^2} \frac{q}{bc}{m_{3/2}}.
\end{align}
The total gaugino masses are then given by
\begin{align}   M_1 &=  \frac{g_Y^2}{16 \pi^2} m_{3/2} \left( 11 \left[ 1 - (\alpha -1)^2 -  \frac{ t(\gamma + t + \gamma t^2)}{1 + \gamma t}  \right] -\alpha (\alpha-1) \frac{6q}{bc}  \right), \notag \\   
M_2 &=  \frac{g_2^2}{16 \pi^2} m_{3/2} \left( \left[1 - 5 (\alpha-1)^2 -5 \frac{ t (\gamma + t + \gamma t^2)}{1 + \gamma t} \right]  +\alpha (\alpha-1) \frac{12q}{bc} \right), \notag \\   
M_3 &= - 3 \frac{g_3^2}{16 \pi^2} m_{3/2} \left[ 1 + (\alpha - 1)^2 + \frac{ t (\gamma + t + \gamma t^2) }{1 + \gamma t} \right]. \label{3B-L:Gauginomasses}   \end{align}  
By using (from eqs.~(\ref{qconstraint}) and (\ref{bc}))
\begin{align}
 \frac{q}{bc} = \frac{q}{q_{\text{max}}} \frac{A(t)(\sigma_s +1) + B(t)}{A(t) \sigma_s + B(t)} \frac{\alpha -1}{2 \alpha},
\end{align}
the corrections to the gaugino masses proportional to $q/q_{\text{max}}$ can be calculated for every $\gamma$. It turns out that these corrections are very small, as can be seen in figure 2, where the gaugino masses are plotted as a function of $q/q_{\text{max}}$ for $\gamma =1.1$.
The low energy spectrum is then expected to be similar to that of the $B-L$ case described above and we therefore do not perform a seperate analysis for the $3B-L$ case.
\begin{figure} \centering \includegraphics[width=0.5\textwidth]{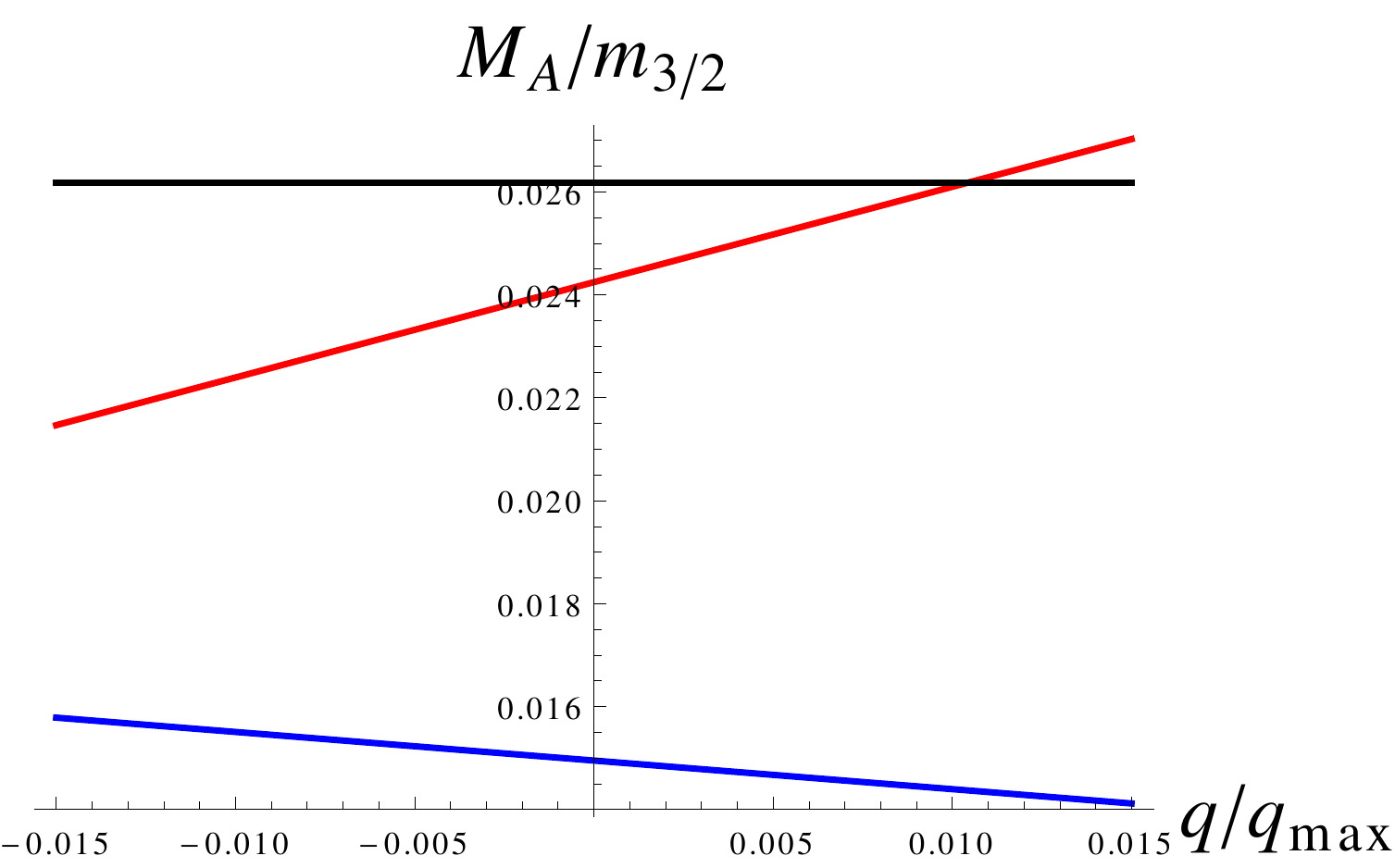} \label{figure2}  
\caption{The gaugino masses $M_1/m_{3/2}$ (blue), $M_2/m_{3/2}$ (red), $M_3/m_{3/2}$ (black) in the $3B-L$ model, as a function of $q/q_{\text{max}}$, where $\frac{5}{3} g_Y^2 = g_2^2 = g_3^2=0.51$ is assumed at the GUT scale. Note that the gaugino masses vary only very little for $\left| q/q_{\text{max}} \right| < 0.013$} 
\end{figure} 

The kinetic terms of the $U(1)_R$ gauge boson are given by\footnote{Alternatively, one can define the gauge kinetic function as $f(s)= 1/q^2$, such that the charge of the fermions is given by (instead of being proportional to) $B-L$.}
\begin{align} \mathcal L_{\text{kin}}/e = -\frac{1}{4} F_{\mu \nu} F^{\mu \nu}.\end{align}
Its mass is given by \cite{RK3}
\begin{align} M_R &= \kappa^{-1} \frac{bc}{\alpha} \notag \\ &= m_{3/2} \left[ (1+\gamma t) e^{\alpha^2 t^2} \sqrt{\frac{\sigma_s A(t) + B(t)}{(\alpha-1)^2}} \right]. \end{align}
In the allowed parameter range this corresponds to $M_R \in \left]25.4 , 99.4 \right[$ TeV. The covariant derivative of the Standard Model fermions $\chi^\alpha$ (with charge $q$) is
\begin{align}  D_\mu \chi^\alpha = \left( \partial_\mu -  iq A_\mu \right) \chi^\alpha, \end{align}
where we have omitted the spin connection and the K\"ahler connection. The charge $q$ of the MSSM fermions satisfies $|q|<0.013  q_{\text{max}} \approx \mathcal O(10^{-17})$.  
We conclude that the $U(1)_R$ gauge boson is (unfortunately) well beyond the current experimental $Z'$ bounds or the corresponding compositeness limits \cite{exp_bounds}.

\section{D-term contributions to the scalar soft masses} \label{sec:other}

In this section we show that another possible solution to the problem of negative soft scalar masses squared in the model proposed in eqs.~(\ref{model:basic}), based on a D-term contribution to the scalar soft masses, does not lead to consistent electroweak vacua. As in the other solutions proposed in \cite{RK3}, solving the problem of tachyonic masses comes at the cost of introducing an extra parameter, $b_1$. In this case the model is given by
\begin{align} \mathcal K &= -\kappa^{-2}\log(s+\bar s) +\kappa^{-2} b(s+ \bar s) + K_{MSSM}, \notag \\  W &= \kappa^{-3} a +  e^{b_1 s} W_{MSSM}, \notag \\  f(s) &= 1 + \beta b s. \label{model:other} \end{align}
Because of the shift symmetry (\ref{shift}), the MSSM superpotential needs to transform (with gauge parameter $\theta$) as
\begin{align}  W_{MSSM} \longrightarrow W_{MSSM} e^{ib_1 c \theta}. \label{WMSSMtransform} \end{align}
The scalar potential is given by
\begin{align}\mathcal V &= V_F + V_D, \notag \\  V_F &= e^K \left[ -3 W \bar W + g^{s \bar s} |\nabla_s W |^2 + |\nabla_\varphi W|^2 \right] \notag \\  V_D &= \frac{1}{2} \left(\kappa^{-2}bc - \frac{\kappa^{-2} c}{s+ \bar s} - q_i \varphi_i \bar \varphi_i \right)^2, \end{align}
where $\varphi_i$ stands for the various MSSM fields and the linear part in the gauge kinetic function has been neglected. Indeed, it is shown in Appendix~\ref{Appendix:nonzerobeta} that $\beta  \ll 1$.

The F-term contribution to the scalar soft masses squared is negative. This can however be compensated by a (positive) contribution proportional to the charge $q_i$ from the D-term scalar potential. This implies that all MSSM fields must have a positive charge under this extra $U(1)$, which is the motivation behind the transformation \ref{WMSSMtransform} and the factor $e^{b_1 s}$ in eq.~(\ref{model:other}). The soft supersymmetry breaking terms can be calculated (with respect to a rescaled superpotential $\hat W = e^{K/2} e^{b_1 s} W$) to be
\begin{align}  m_0^2 &= m_{3/2}^2 (\sigma_s + 1) +\kappa^{-2} q_i bc \frac{ 1 - \alpha}{\alpha}, \notag \\  A_0 &= m_{3/2} \ \rho_s,  \notag \\  B_0 &= m_{3/2} \ (\rho_s +1), && \rho_s &= -1 + (\alpha-1)(\alpha - 1 + b_1 (s + \bar s) ), \end{align}
where $\alpha$ and $\sigma_s$ are defined in eqs.~(\ref{model:tuning}). The gravitino mass is given by 
\begin{align} m_{3/2} &=  \kappa^{-1} a e^{\alpha/2} \sqrt{\frac{b}{\alpha}}. \label{p1:gravitino_mass} \end{align}
The relations (\ref{model:tuning}) can be rewritten as
 \begin{align} bc = - \kappa m_{3/2} e^{-\alpha/2} \sqrt{\alpha A(\alpha)}, \label{bc=alphaA} \end{align}
 which can be used to rewrite the D-term contribution to the mass in the form 
\begin{align}  m_0^2 &= m_{3/2}^2 (\sigma_s + 1) + m_{3/2} \kappa^{-1} q_i \ Q(\alpha), \notag \\  Q(\alpha) &= \frac{\alpha-1}{ \alpha} e^{-\alpha/2} \sqrt{\alpha A(\alpha)} \approx 0.8598. \label{Qalpha} \end{align} 
To avoid tachyonic masses, the charges of the MSSM fields should satisfy
\begin{align} q_i > q_0 = - \kappa  m_{3/2} \frac{\sigma_s +1}{Q(\alpha)} \approx \  0.558 \ \kappa  m_{3/2},   \label{qlimit} \end{align}
which corresponds to $q \gtrsim 0.5 \times 10^{-15}$ for $m_{3/2} \approx 20$ TeV. 
The (non-universal) scalar soft masses and trilinear terms can be summarized as
\begin{align}
 m_{0,i}^2 &= m_{3/2}^2 \left[ (\sigma_s +1) \left( 1 - \frac{q_i}{q_0} \right) \right], \notag \\
 A_0 &= m_{3/2} \left[ -1 + (\alpha -1)\left(\alpha -1 + \frac{q}{q_0} P(\alpha)\right) \right], \notag \\
 P(\alpha) &=   \frac{2 \alpha e^{\alpha/2} (\sigma_s +1)}{\sqrt{\alpha A(\alpha)} Q(\alpha)} \approx 0.799,
\end{align}
where eqs.~(\ref{model:tuning}) were used and $q = b_1 c /2$. The charges $q_i$ are given in terms of three independent parameters $\theta$, $\theta_Q$ and $\theta_L$ by
\begin{align} q_{H_u} &= \theta q, \notag \\ q_{H_d} &= (2-\theta)q, \notag \\ q_L &= \theta_L q, \notag \\ q_{\bar e} &= (\theta - \theta_L) q, \notag \\
q_{Q} &= \theta_Q q, \notag \\q_{\bar u} &= (2-\theta - \theta_Q)q, \notag \\ q_{\bar d} &= (\theta - \theta_Q)q, \label{charges}\end{align}
such that 
\begin{align}  q_{H_u} + q_{H_d} &= 2q, \notag \\  q_{\bar e} + q_{L}+ q_{H_d} &= 2q, \notag \\  q_{\bar d} + q_Q + q_{H_d} &= 2q, \notag \\  q_{\bar u} + q_Q + q_{H_u} &= 2q,\end{align}
and eq.~(\ref{WMSSMtransform}) is satisfied for the MSSM superpotential (\ref{MSSMsuperpot}).

Next, the cubic and mixed anomalies are proportional to
\begin{align} \mathcal C_R &= q^3 f(\theta,\theta_L,\theta_Q), \notag \\  \mathcal C_1 &=  \frac{3q}{2} \left(6- 3 \theta_Q - \theta_L  \right), \notag \\
 \mathcal C_2 &= q \left(2+ 3 \theta_L + 9 \theta_Q \right), \notag \\ \mathcal C_3 &= 6q,\end{align}
where
\begin{align} f(\theta,\theta_L,\theta_Q) &=  3 \left( 6 \theta_Q^3 + 3(2-\theta_Q - \theta_L)^3+ 3(\theta-\theta_Q)^3 + 2 \theta_L^3 + (\theta- \theta_L)^3 \right) \notag \\
& \qquad + 2 \left( (2-\theta)^3 + \theta^3 \right). \label{ftheta}\end{align}
The Green-Schwarz counter terms are then proportional to (see Appendix \ref{appendix:anomalies})
\begin{align}
 \beta &= - q^3 \frac{f(\theta,\theta_L,\theta_Q) }{12 \pi^2 b c}, \notag \\  \beta_1 &= - q\frac{  3 \left(6- 3 \theta_Q - \theta_L  \right)}{8 \pi^2 c}, \notag \\
 \beta_2 &= - q\frac{ \left(2+ 3 \theta_L + 9 \theta_Q \right)}{4 \pi^2 c}, \notag \\ \beta_3 &= - q\frac{3}{2 \pi^2 c}. \label{other:anomalycancellation}
\end{align}
This results in contributions to the gaugino masses
\begin{align} M_A &= -\frac{g_A^2}{2} \frac{\alpha (\alpha - 1)}{b} m_{3/2} \beta_A. \end{align}
From eqs.~(\ref{model:tuning})  the Green-Schwarz contributions to the gaugino masses are given by
\begin{align}
 M_1 &= - \frac{g_Y^2}{16\pi^2} 3(6-3\theta_Q-\theta_L) \frac{ \alpha (\alpha - 1) e^\alpha}{\sqrt{\alpha A(\alpha)} } \kappa^{-1}q, \notag \\
 M_2 &= - \frac{g_2^2}{16\pi^2}2 (3 + \theta_L + 9 \theta_Q) \frac{2 \alpha (\alpha - 1) e^\alpha}{\sqrt{\alpha A(\alpha)}} \kappa^{-1}q, \notag \\
 M_3 &= - \frac{g_3^2}{16\pi^2} \frac{ 12 \alpha (\alpha - 1) e^\alpha}{\sqrt{\alpha A(\alpha)}} \kappa^{-1}q.
\end{align}

The anomaly mediated contribution to the gaugino masses are given by eq.~(\ref{Appendix:gauginomassAM}). The total one-loop gaugino mass is the sum of these contributions 
\begin{align}
 M_1 &= - \frac{g_Y^2}{16\pi^2}  \left(  11 m_{3/2} \left[-1 + (\alpha -1)^2 \right] -  \frac{3}{2} (6-3\theta_Q-\theta_L) \frac{ (\alpha - 1)^2}{Q(\alpha) } \kappa^{-1}q \right), \notag \\
 M_2 &= - \frac{g_2^2}{16\pi^2} \left( m_{3/2} \left[ -1 + 4 (\alpha - 1)^2 \right] - (3 + \theta_L + 9 \theta_Q)    \frac{ (\alpha - 1)^2}{Q(\alpha) } \kappa^{-1}q \right), \notag \\
 M_3 &= - \frac{g_3^2}{16\pi^2} \left( 3 m_{3/2} \left[ 1+ (\alpha-1)^2\right] -6  \frac{ (\alpha - 1)^2}{Q(\alpha) } \kappa^{-1}q \right).
\end{align}
The soft terms can be summarized as (where $\xi= q/q_0 > 2$, and eqs.~(\ref{Qalpha}) and (\ref{bc=alphaA}) were used to rewrite the gaugino masses)
\begin{align}
 m_{0,i}^2 &= m_{3/2} \left[ (\sigma_s +1) \left( 1 - \theta_i \xi \right) \right], \notag \\
 A_0 &= m_{3/2} \left[ -1 + (\alpha -1)\left(\alpha -1 + \xi P(\alpha)\right) \right], \notag \\
 B_0 &= A_0 + m_{3/2}, \notag \\
M_1 &= - \frac{g_Y^2}{16\pi^2} m_{3/2}  \left(  11 \left[-1 + (\alpha -1)^2 \right] +   \frac{3}{2} (6-3\theta_Q-\theta_L)  \frac{ (\alpha - 1)^2}{Q(\alpha)^2 }(\sigma_s +1)  \right), \notag \\
 M_2 &= - \frac{g_2^2}{16\pi^2}  m_{3/2} \left(  \left[ -1 + 4 (\alpha - 1)^2 \right] +  (3 + \theta_L + 9 \theta_Q)  \frac{ (\alpha - 1)^2}{Q(\alpha)^2 } (\sigma_s +1) \right), \notag \\
 M_3 &= - \frac{g_3^2}{16\pi^2}  m_{3/2} \left( 3  \left[ 1+ (\alpha-1)^2\right] +6  \frac{ (\alpha - 1)^2}{Q(\alpha)^2 } (\sigma_s +1) \right).
\end{align}
The above soft terms depend on five parameters, namely $m_{3/2}$, $\xi$, $\theta$, $\theta_L$ and $\theta_Q$. Following, a parameter scan has been performed \cite{softsusy} for $m_{3/2} \in [15 \text{ TeV},40 \text{ TeV}]$, $\xi \in [2,10]$, $\theta$, $\theta_L$, $\theta_Q$ $\in ]0,2[$, tan$\beta \in [1,60]$. For $m_{3/2} < 15$ TeV, the gaugino masses are expected to be experimentally excluded, $\xi > 2$ in order to satisfy the constraint (\ref{qlimit}). $A_0$ is negative and monotonically decreasing with $\xi$, such that for $\xi >10$ the trilinear term becomes $A_0 < - 9 m_{3/2}$. In principle, the value of tan$\beta$ (which is the ratio between the two Higgs VEVs) is fixed by $B_0$ \cite{Ellis-B-tanbeta}, however we performed a scan over all possible values of tan$\beta$ instead. Such a high value for $|A_0|$ would contribute to the RGE for the stop mass parameter, so that it runs to a negative value before the electroweak symmetry breaking scale is reached.\footnote{Or the stau becomes tachyonic for a very small region of the parameter space.}

In this parameter range, no viable electroweak symmetry breaking conditions were found. We conclude that, even though the above idea is very appealing from a theoretical point of view, one cannot (at least in this model) use a D-term contribution to the scalar soft masses proportional to the charge of a MSSM field under an extra $U(1)$ factor to solve the problem with tachyonic masses.

\section{Conclusions}

In this work we studied a simple model~\cite{RK3,Zwirner,RK1} of supersymmetry breaking in supergravity based on a single chiral multiplet and a gauged shift symmetry that we identify with a known global symmetry of the Standard Model. The model allows for a tiny and tunable cosmological constant while leaving the gravitino mass (and thus the supersymmetry breaking scale) separately free. We analyzed the phenomenological implications in great detail for the particular case where the global symmetry is $B-L$, or $3B-L$, which contains the known matter parity of the MSSM as a subgroup. 
The latter combination has also the advantage of forbidding all dimension-four and dimension-five operators violating baryon or lepton number in the MSSM. We showed that the phenomenology is similar to the one obtained in~\cite{RK3}, where the MSSM fields are inert under the shift symmetry, with the exception of the stop mass which can be become lighter to about $1.5$ TeV (compared to $2$ TeV).

The above model contains in its (hidden) supersymmetry breaking sector an extra Polonyi-like field, bringing an additional parameter, to avoid the appearance of tachyonic scalar soft masses for the MSSM fields when these are inert under the shift symmetry~\cite{RK3}. Alternatively, one can add non-canonical kinetic terms for the MSSM fields bringing again a parameter with a similar phenomenology. 
We showed that the problem of tachyonic scalar masses cannot be solved from the presence of the (positive) D-term contributions to the scalar potential, proportional to the charge of the MSSM superfields under the shift symmetry, because of incompatibility with the existence of a viable electroweak vacuum.
Finally, we explored (in an Appendix) another possibility that was left open in our previous works where one has a linear gauge kinetic term and the coefficient of the logarithm in the dilaton K\"ahler potential is $p=2$ (instead of $p=1$), and showed that it does not work either because it leads to an unacceptable high value of the gravitino mass.

\section*{Acknowledgements}
This research was (partly) supported by the NCCR SwissMAP, funded by the Swiss National Science Foundation. R.K. would like to thank S.~Richter for useful discussions.

\appendix

\section{Anomaly Cancellation} \label{appendix:anomalies}

Anomalies in supergravity theories with Fayet-Iliopoulos terms are treated in full detail in \cite{FreedmanAnomalies,RK2}, we here summarize their results and apply them to our model. The cubic $U(1)_R$ anomaly, proportional to $\mathcal C_R= \Tr[Q^3]$, gives the following anomalous contribution (with gauge parameter $\theta$) to the Lagrangian 
\begin{align} \delta \mathcal L_{1-loop} &=  - \frac{\theta}{32 \pi^2} \  \frac{ \mathcal C_R}{3} \ \epsilon^{\mu \nu \rho \sigma} F_{\mu \nu} F_{\rho \sigma}, \label{1-loop} \end{align}
which is canceled by a Green-Schwarz mechanism as follows: If the gauge kinetic function contains a linear contribution in $s$, say $f(s) = 1/g^2 + \beta_R s$, the Lagrangian contains a contribution
\begin{align} \mathcal L_{GS} = \frac{1}{8}\text{Im}\left(f(s) \right) \ \epsilon^{\mu \nu \rho \sigma} F_{\mu \nu} F_{\rho \sigma}. \end{align}
The gauge variation of this Green-Schwarz contribution,
\begin{align} \delta \mathcal L_{GS} &= -\frac{\theta \beta_R c}{8} \ \epsilon^{\mu \nu \rho \sigma} F_{\mu \nu} F_{\rho \sigma}  \label{Green-Schwarz_variation}\end{align}
cancels the cubic anomaly (\ref{1-loop}), provided 
\begin{align} \beta_R = - \frac{ \mathcal C_R}{12 \pi^2 c}. \end{align}
Similarly, the mixed anomalies with the Standard Model gauge groups, proportional to $\mathcal C_A$, are canceled (with the inclusion of appropriate generalized Chern-Simons terms \cite{GCS}), provided
  \begin{align}   \mathcal C_A &= - 4 \pi^2 c \ \beta_A , \ \ \ \ \ \ \ A=Y,2,3 .  \end{align}

Since the gaugino masses are proportional to derivatives of their respective gauge kinetic functions   
\begin{align}
M_A &= -\frac{g_A^2}{2} e^{\kappa^2 K/2} \partial_\alpha f_{A}(S) g^{\alpha \bar \beta} \bar \nabla_{\bar \beta} \bar W \notag \\
&=-\frac{g_A^2}{2} e^{\kappa^2 K/2} \beta_A g^{s \bar s} \bar \nabla_{\bar s} \bar W , \label{gauginomassGS}
\end{align} an 'anomalous' $U(1)_R$ can give a contribution (at one-loop) to the mass of the gauginos.

It should however be noted that at one-loop there is another contribution to the gaugino masses, due to a mechanism called anomaly mediation \cite{gauginomass}, given by
\begin{align} M_A = - \frac{g_A^2}{16 \pi^2} \left[ (3 T_G - T_R) m_{3/2} + (T_G - T_R) \mathcal K_\alpha F^\alpha + 2 \frac{T_R}{d_R} (\log \det \mathcal K|_R \ '')_{,\alpha} F^\alpha \right] , \label{Appendix:gauginomassAM}  \end{align}
 where $T_G$ is the Dynkin index of the adjoint representation, normalized to $N$ for $SU(N)$, and $T_R$ is the Dynkin index associated with the representation $R$ of dimension $d_R$,   equal to $1/2$ for the $SU(N)$ fundamental\footnote{For $U(1)_Y$ we have $T_G = 0$ and $T_R = 11$, for $SU(2)$ we have $T_G = 2$ and $T_R = 7$, and for $SU(3)$ we have $T_G = 3$ and $T_R = 6$.}. An implicit sum over all matter representations is understood. The quantity $3T_G - T_R$ is the one-loop beta function coefficient.   The expectation value of the auxiliary field $F^\alpha$, evaluated in the Einstein frame is given by   
  \begin{align}   F^\alpha = - e^{ \kappa^2 \mathcal K/2} g^{\alpha \bar \beta} \bar \nabla_{\bar \beta} \bar W.   \end{align}
It was shown by the authors in \cite{RK3} that indeed both contributions of eqs.~(\ref{gauginomassGS}) and (\ref{Appendix:gauginomassAM}) to the gaugino mass should be taken into account. 

\section{On the consistency of $p=2$} \label{Appendix:p2}

In this Appendix we focus on the case $p=2$ 
\begin{align} \mathcal K &= - \kappa^{-2}p\log(s + \bar s) + \kappa^{-2} b(s + \bar s) + K(z,\bar z), \notag \\  W &= \kappa^{-3} a ,\end{align}
where an extra chiral multiplet $z$, which can either be a hidden sector field or a MSSM field, has been added to the model. Since $p=2$, the tunability of the scalar potential (in order to allow for a tunable small cosmological constant) requires the gauge kinetic function to be linear in $s$~\cite{RK3}
\begin{align} f(s) = \beta s. \end{align}
This results in a Green-Schwarz contribution (\ref{Green-Schwarz_variation}) which would make the Lagrangian non-gauge invariant. The extra field $z$ has a charge $q$ under the extra $U(1)$ and is added such that its contribution to the cubic anomaly cancels the Green-Schwarz contribution above. The scalar potential is then given by
\begin{align}  \mathcal V &= V_F + V_D, \notag \\  V_F &= \kappa^{-4} a^2 e^{\mathcal K} \left( \sigma_s + \kappa^2 z \bar z \right), 		&\sigma_s = -3 + \frac{(s + \bar s)^2}{2} \left( b - \frac{2}{s + \bar s} \right)^2, \notag \\  V_D &= \frac{1}{2\beta (s+ \bar s)}\left( \kappa^{-2}c \left(b - \frac{2}{s + \bar s} \right) - q z \bar z\right)^2. \end{align}
The mass of the field $z$ is given by 
\begin{align}  m_z^2 = m_{3/2}^2 (\sigma_s + 1) +  \kappa^{-2} \frac{ b^2 c q }{\beta} \frac{2-\alpha}{\alpha^2}, \end{align}
in the notation of \cite{RK1}, where $\langle s + \bar s \rangle b= \alpha \approx -0.1832$. The relation between the parameters which enforces a vanishing cosmological constant is given by
\begin{align}  \frac{\beta a^2}{bc^2} = A(\alpha) \approx -50.66, \label{p2:Aalpha} \end{align}
 and the anomaly cancellation condition is
\begin{align} \beta = - \frac{q^3}{12 \pi^2 c}. \label{p2:anomalycancellation}\end{align}
This fixes the sign of $q$ and results in a negative D-term contribution to the scalar mass squared of $z$, which by using eqs.~(\ref{p2:Aalpha}) and (\ref{p2:anomalycancellation}) is
\begin{align}  m_z^2 &= m_{3/2}^2 (\sigma_s + 1) -  \kappa^{-2/3} m_{3/2}^{\ 4/3} R(\alpha), \notag \\
R(\alpha) &= \left( \frac{12 \pi^2 (2-\alpha)^3 e^{-2 \alpha} }{A(\alpha) \alpha^2} \right)^{1/3} \approx  2.74. \end{align}
The constraint that the mass of $z$ remains non-tachyonic 
\begin{align}  m_{3/2}^2 (\sigma_s + 1) > \kappa^{-2/3} m_{3/2}^{\ 4/3} R(\alpha), \end{align}
forces the mass of the gravitino to exceed the Planck scale 
\begin{align} m_{3/2} > \frac{R(\alpha)}{\sigma_s +1} \kappa^{-1} \approx 7.16 \ \kappa^{-1}.\end{align}
One concludes that one cannot use the parameter $p$ to solve the problem of tachyonic masses outlined in section \ref{sec:model}.
The above argument can easily be generalized to include several charged fields  (unless their number becomes extremely large).

\section{Comments on nonzero $\beta$} \label{Appendix:nonzerobeta}

In this Appendix we first demonstrate how tunable dS vacua can be found for the model~(\ref{model:other}) for a general $\beta$.
In section~\ref{sec:other} it was assumed that $\beta \ll 1$. Here, we show that the anomaly cancellation conditions indeed force $\beta$ to be very small.
 
\begin{figure} \centering \includegraphics[width=0.5\textwidth]{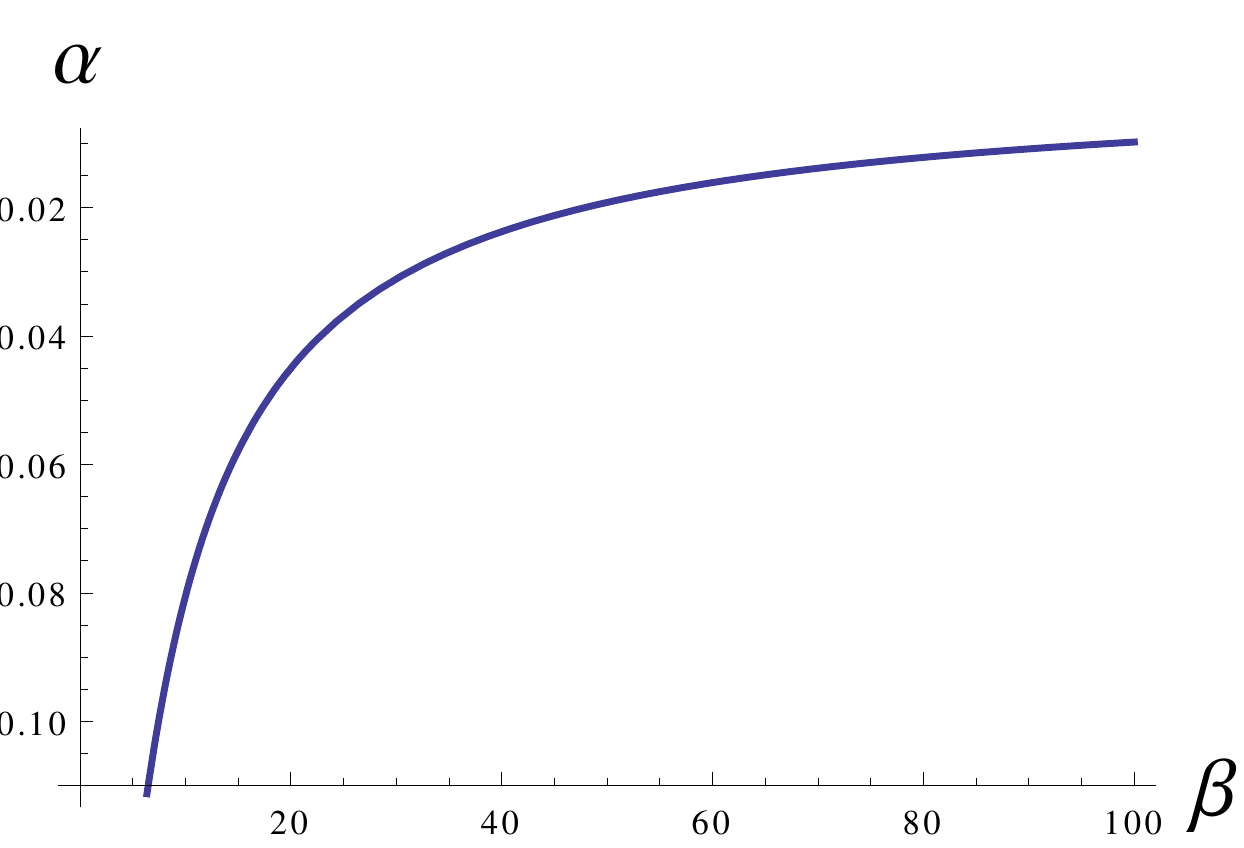} 
 \caption{The solutions to eq.~(\ref{betaprime}) are plotted with $\alpha$ as a function of $\beta$. Although up to four (complex) solutions can exist for every $\beta$, only a single solution satisfies $\alpha < 0$ (to ensure a negative $b$) and $A(\alpha,\beta) < 0$ to satisfy eq.~(\ref{Aalphabeta}). Only positive $\beta$ is plotted since the positivity is required to satisfy the anomaly cancellation conditions (see below). }  \label{plot_alphabeta} \end{figure}

If $\beta$ is not neglected, the D-term contribution to the scalar potential is given by
\begin{align} \mathcal V_D =  \frac{1}{2 + \beta b (s + \bar s)} \left(\kappa^{-2}bc - \frac{\kappa^{-2} c}{s+ \bar s} - q_i \varphi_i \bar \varphi_i \right)^2.\end{align}
The solution to $\partial \mathcal V = \mathcal V = 0$ gives 
\begin{align} (\alpha - 1)^2 \left( \alpha^2 -2 \right) + \left( \frac{\beta \alpha (\alpha-1)}{2+\beta \alpha} -2 \right) \left(-3 + (\alpha-1)^2 \right)=0. \label{betaprime}\end{align}
For $\beta = 0$, this gives indeed $\alpha \approx -0.233$. For other values of $\beta$ the result is plotted in figure \ref{plot_alphabeta}.
The relation between the parameters to obtain a vanishing cosmological constant becomes
\begin{align} \frac{bc^2}{a^2} = - \frac{\alpha e^{\alpha} (2+ \beta \alpha) (-3 + (\alpha-1)^2 )}{(\alpha-1)^2} =A(\alpha,\beta). \label{Aalphabeta}\end{align}
One concludes that for every (finite) $\beta$ the vacuum is tunable. The gravitino mass eq.~(\ref{p1:gravitino_mass}) is repeated here for convenience of the reader
\begin{align} m_{3/2} &=  \kappa^{-1} a e^{\alpha/2} \sqrt{\frac{b}{\alpha}}. \notag \end{align}
The scalar soft mass squared and the trilinear couplings are given by
\begin{align}  m_0^2 &= m_{3/2}^2 (\sigma_s + 1) +\kappa^{-2} \frac{q_i bc}{1+ \beta \alpha/2} \frac{ 1 - \alpha}{\alpha}, \notag \\ 
 A_0 &= m_{3/2} \ \rho_s,  \notag \\ 
 \rho_s &= -1 + (\alpha-1)\left(\alpha - 1 + P(\alpha,\beta) \xi \right), \notag \\
 P(\alpha,\beta) &= \frac{2 \alpha e^\alpha (\sigma_s +1) }{(\alpha-1) A(\alpha,\beta)}  \left(1+\frac{ \beta \alpha}{2}\right). \label{Palphabeta} \end{align}
The relations~(\ref{Aalphabeta}) can be written as
\begin{align} bc = -\kappa m_{3/2} e^{-\alpha/2} \sqrt{\alpha A(\alpha, \beta)}, \label{bc=Aalphabeta}\end{align}
which are used to write the scalar soft mass squared as 
\begin{align}  m_0^2 &= m_{3/2}^2 (\sigma_s + 1) + m_{3/2} \kappa^{-1} q_i \ Q(\alpha,\beta), \notag \\  
Q(\alpha, \beta) &= \frac{\alpha-1}{ \alpha} e^{-\alpha/2} \frac{\sqrt{\alpha A(\alpha,\beta)} }{1+\beta \alpha/2}.  \end{align}
To avoid tachyonic masses, the charges of the MSSM fields should satisfy
\begin{align} q_i > q_0 = - \kappa  m_{3/2} \frac{\sigma_s +1}{Q(\alpha,\beta )}. \label{qlimitbeta} \end{align}

\begin{figure}[H] \centering \includegraphics[width=0.5\textwidth]{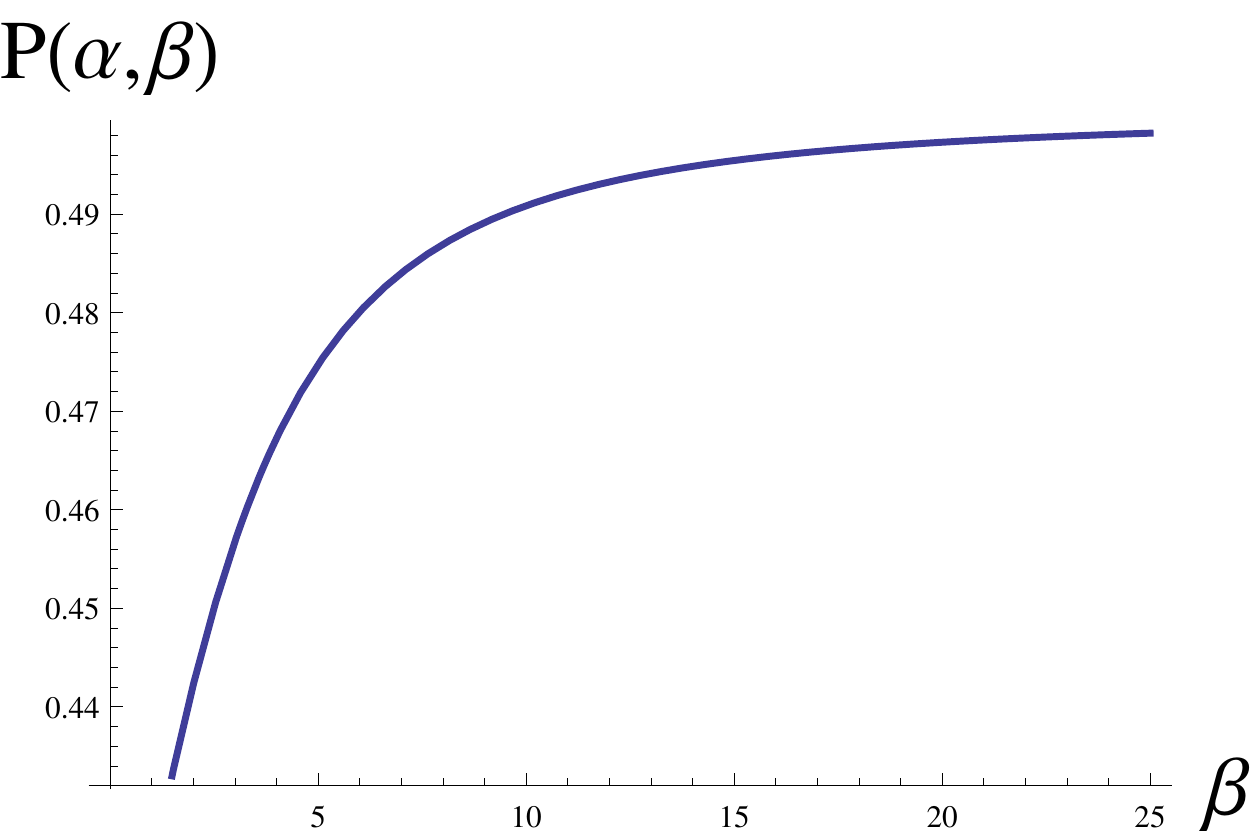} 
\caption{A plot of $P(\alpha,\beta)$ given by eq.~(\ref{Palphabeta}) as a function of $\beta$, where $\alpha$ and $\beta$ are solutions of eq.~(\ref{betaprime}). It follows that the trilinear coupling $A_0$ (as a function of $\xi$) has only a very small dependence on $\beta$. One should therefore focus on $\xi \lesssim 10$ (as in section \ref{sec:other}) since otherwise $|A_0|$ becomes too large to allow for a viable electroweak vacuum.}  \label{plot_Pbeta} \end{figure}

Next, the anomaly cancellation conditions (see eqs.~(\ref{other:anomalycancellation})) for the cubic anomaly give
\begin{align} \beta &= - \frac{f(\theta,\theta_L,\theta_Q) q^3}{12 \pi^2 bc} \notag \\
 &= f(\theta,\theta_L,\theta_Q) \xi^3 (\kappa m_{3/2})^2 g(\alpha,\beta), \label{betaprimeanomaly}\end{align}
where eqs.~(\ref{bc=Aalphabeta}) and (\ref{qlimitbeta}) were used, and
\begin{align} g(\alpha,\beta ) = - \frac{e^{\alpha/2} (\sigma_s +1)^3}{12 \pi^2 Q(\alpha,\beta )^3 \sqrt{\alpha A(\alpha,\beta )}}. \label{galphabeta}\end{align}
From eq.~(\ref{ftheta}) it follows that $ 9.47 \leq f(\theta,\theta_L,\theta_Q) \leq 91$, for all $\theta,  \theta_L,  \theta_Q  \in \  ]0,1[$, such that $\beta >0$  from eq.~(\ref{betaprimeanomaly}).
From figure~\ref{plot_Pbeta}, where $P(\alpha,\beta)$ is plotted as a function of $\beta$, it follows that the trilinear coupling $A_0$ has only a very slight dependence on $\beta $. 
We can therefore assume $\xi < \mathcal O(10)$ (as in section~\ref{sec:other})  since otherwise the trilinear coupling $A_0$ would be too large to allow for a realistic electroweak vacuum.
\begin{figure}[H] \centering \includegraphics[width=0.5\textwidth]{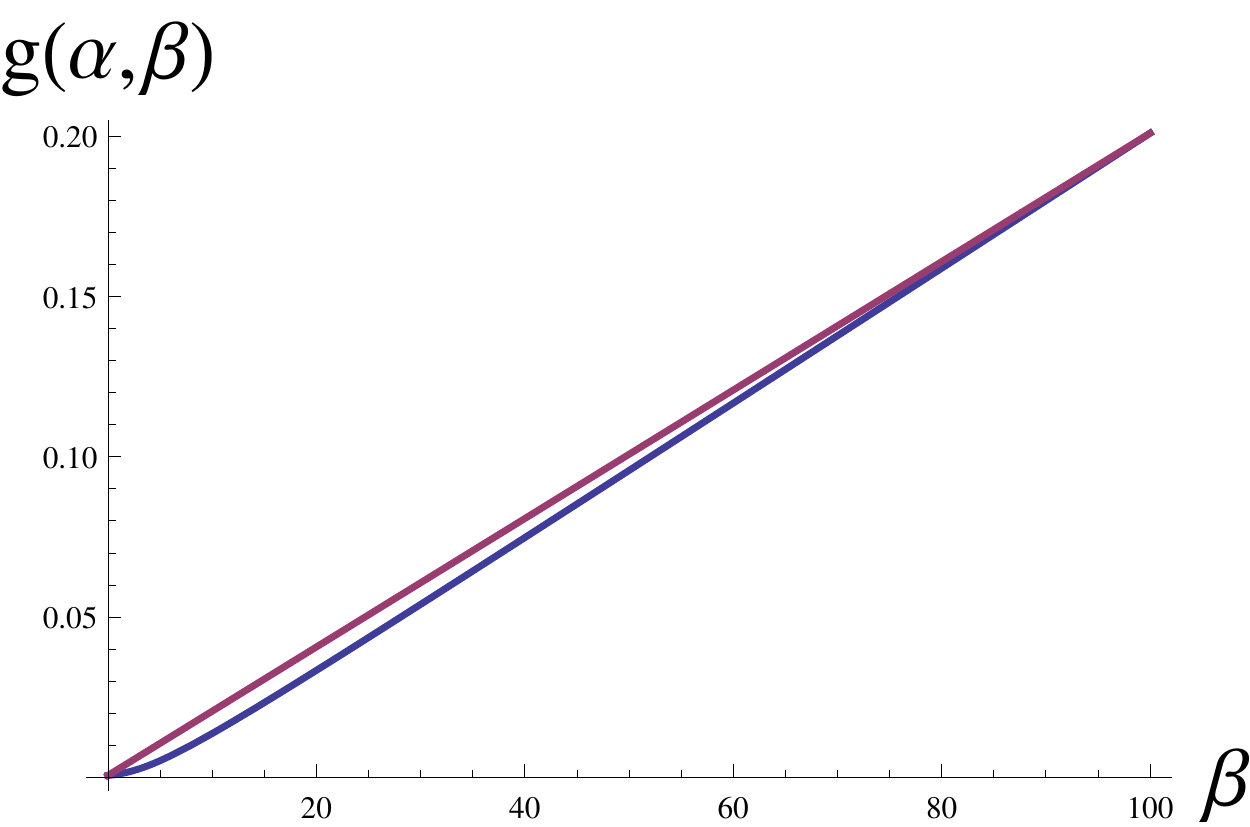} 
\caption{In blue: A plot of $g(\alpha,\beta)$ given by eq.~(\ref{galphabeta}) as a function of $\beta$, where $\alpha$ and $\beta$ are solutions of eq.~(\ref{betaprime}). The approximation $g_0(\beta)$ given by eq.~(\ref{g0}) is plotted in purple.}
  \label{plot_gbeta} \end{figure} 

In figure \ref{plot_gbeta},  $g(\alpha,\beta)$ is plotted as a function of $\beta $. As is shown, we can approximate $g(\alpha,\beta )$ by a linear function of $\beta$ (In fact, $g_0(\beta)$ is also plotted in figure~\ref{plot_gbeta} and completely overlaps with the actual function $g(\alpha,\beta)$)
\begin{align} g(\alpha,\beta) \approx g_0(\beta) &= g(0) + \omega \ \beta,\label{g0} \end{align} 
where \begin{align}g(0) &= 0.0005618, \notag \\ \omega &= 0.002003. \end{align}
For a given $m_{3/2}$, $\xi$ and $f(\theta,\theta_L,\theta_Q)$, one can then solve $\beta = f(\theta,\theta_L,\theta_Q) \xi^3 (\kappa m_{3/2})^2 g_0(\beta)$ to find
\begin{align} \beta = \frac{g(0)}{1/\mathcal A - \omega}, \label{betaapprox} \end{align}
where \begin{align}\mathcal A =  f(\theta,\theta_L,\theta_Q) \xi^3 (\kappa m_{3/2})^2.   \end{align}
For $\xi < 10$, $m_{3/2} <40 \text{ TeV}$ and $f(\theta,\theta_L,\theta_Q) <91$, this corresponds to $\beta \lesssim  \mathcal O(10^{-26}) $. Note that for small $\beta$ one can approximate eq.~(\ref{betaapprox}) by $\beta \lesssim g(0) \mathcal A$ which leads to the same result.

We conclude that the anomaly cancellation condition can only be satisfied for very small $\beta$.

\vfill
\end{document}